\documentclass[aps,amssymb,twocolumn,prb,floatfix,showpacs,a4paper]{revtex4}

\usepackage[dvips]{graphicx}
\usepackage[hypertex,breaklinks=true,colorlinks=false]{hyperref}
\begin{document}
\title{
Specific  heat of the S=$\frac{1}{2}$ Heisenberg   model on the kagome
lattice: high-temperature series expansion analysis}

\author{G.  Misguich}
\email{misguich@spht.saclay.cea.fr}
\affiliation{Service de Physique Th{\'e}orique,  URA 2306 of CNRS\\
CEA Saclay, 91191 Gif-sur-Yvette C\'edex, FRANCE}
\author{B.  Bernu}
\email{bernu@lptl.jussieu.fr} \affiliation{ Laboratoire  de Physique
Th{\'e}orique   des Liquides,
Universit{\'e}  P. et M. Curie, URA 765 of  CNRS\\
Case 121, 4 Place Jussieu, 75252 Paris C\'edex 05, FRANCE.}
\begin{abstract}

We compute  specific heat of  the antiferromagnetic spin-$\frac{1}{2}$
Heisenberg model on the kagome lattice.   We use a recently introduced
technique  to analyze high-temperature series  expansion  based on the
knowledge of high-temperature series  expansions, the total entropy of
the system  and the low-temperature expected  behavior of the specific
heat  as   well   as the    ground-state energy.    In  the  case   of
kagome-lattice antiferromagnet, this method predicts a low-temperature
peak at $T/J\lesssim 0.1$.

\end{abstract}

\maketitle
\section{Introduction}

We consider  the    nearest-neighbor Heisenberg model   on the  kagome
lattice:
\begin{equation}
    \mathcal{H}= 2\sum_{\left<i,j\right>} \vec{S}_i\cdot\vec{S}_j
    \label{eq:H}
\end{equation}
Because of its    unconventional properties,  the   spin-$\frac{1}{2}$
kagome antiferromagnet (KAF) has  been subject to an intense  activity
these  last     years.   All studies     agree   that  this frustrated
two-dimensional  magnet has  no   long-ranged magnetic order  at  zero
temperature.\cite{ze90,ce92,sh92,le93,ze95,nm95,lblps97,yf00,bcl02}  Exact
diagonalization studies have  established that the low-energy spectrum
of the kagome-lattice Heisenberg antiferromagnet has a large number of
spin-singlet states before the first spin 1 excited state.\cite{web98}
Among the different theories developed  to explain this unconventional
spectrum, short-range Resonating Valence-Bond (RVB) pictures have been
proposed.\cite{m98,mm00,msp03,ns03,ba04}

The high-temperature (HT)  expansion  of  the  specific heat as   been
computed up to order $1/T^{16}$ by  Elstner and Young.\cite{ey94}
We have checked and  extended this series to  order $1/T^{17}$.  The
additional    term  for the  specific   heat  per  site  is  given  by
$c_v(T)=\frac{3}{2}\beta^2+\cdots+\frac{1845286680253}{366912000}\beta^{17}$. Elstner
and    Young  analyzed  the    series  through  conventional  Pad{\'e}
approximants with the additional  constraint that the specific heat
must vanish at  $T=0$.  At the  highest orders, they found a  specific
heat curve  with  a single  maximum  around $T=1.3$  but  with a large
entropy     deficit       of                about                40\%:
$\int_0^\infty{c_v(T)/TdT}\simeq0.6\ln(2)$.     They  concluded    the
existence of a low-temperature  structure corresponding to an  entropy
of  about 40{\%} of $\ln2$ and  claimed that this low-energy structure
could   not be  accessed from  the   high-temperature expansion of the
specific heat.   They   argued that even  though  the  {\em classical}
kagome  antiferromagnet has   a  non-vanishing ground-state   entropy,
quantum fluctuations in the  spin-$\frac{1}{2}$ model are expected  to
lift this degeneracy.

In this paper, we  revisit the question of the  specific heat with the
help of a new method to analyze high-temperature series data. Compared
to the usual Pad\'e approximant approach, this method\cite{bm01} takes
advantage  of additional information on  the system: the two sum rules
on  the energy  and on  the  entropy  are exactly  satisfied.  In many
simple systems (one- and two  dimensional ferro- or antiferromagnets),
this technique   allows to compute  accurately  the specific heat {\em
down to zero temperature},\cite{bm01}  which is  not  the case  if one
does a direct Pad\'e  analysis of the series.   For the present kagome
model we show   that   this method   provides  rich  semi-quantitative
informations  on the specific heat  curve, although a full convergence
down to zero temperature cannot be achieved.

\section{Direct high-temperature expansion of the specific heat}

\begin{figure}
\begin{center}
\includegraphics[width=5cm]{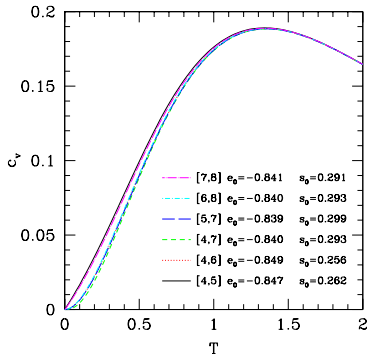}
\end{center}
\caption{(color online) Specific  heat    obtained  from    Pad\'e  approximants
to  the high-temperature series  of $c_v(T)$.   Only approximations of
degrees   $[u,u+1]$,    (resp.     $[u,u+2]$   and    $[u,u+3]$)   are
considered. They  vanish at zero temperature as  $T$ (resp.  $T^2$ and
$T^3$).  The only $6$ such approximants  from order 9  to 17 which are
positive on the positive real  axis are shown. The ground-state energy
per site $e_0$ and entropy $s_0$ obtained  by integrating these Pad\'e
approximations are indicated.}
\label{fig:PadeCV}
\end{figure}

We reproduce here the first attempt by  Elstner and Young\cite{ey94} to
compute the specific  heat from its high-temperature  expansion alone.
We use  Pad\'e approximants to extrapolate the  series.  We impose the
specific heat to vanish at low temperature as $T$, $T^2$ or $T^3$.  At
orders $9$ to $17$, only $6$ such approximants do not develop poles or zeros
in the  interval $T\in]0,\infty]$ (Fig.~\ref{fig:PadeCV}).  One should
notice  at this point  that   the remaining Pad\'e approximants  agree
reasonably  well down  to zero  temperature.   This is usually not the
case in   2D  antiferromagnet where even   the  position of   the peak
($T\simeq 1$) can  hardly be  obtained by  the use   of direct Pad\'e
approximants to  the series for the   specific heat.\cite{bm01}  From
this point of view, the HT series  expansion of the kagome model seems
to  have   a    faster convergence     than    models  such  as    the
triangular-lattice antiferromagnet.

By  integration of these  approximants,  we evaluate the  ground-state
energy $e_0=\int_0^\infty   c_v(T)dT$  and  the   ground-state entropy
$s_0=\log(2)-\int_0^\infty c_v(T)dT$.   These values  are indicated in
Fig.~\ref{fig:PadeCV}.  The   ground-state energy is  about  $-0.845$,
only slightly higher  ($0.02$) than  estimations obtained  from  exact
diagonalizations.  The entropy deficit  is very large: $0.3$ ($40${\%}
of  $\log(2)$).   Elstner     and Young\cite{ey94}    argued  that   a
low-temperature  peak should be present in  the specific heat in order
to compensate  the deficit of $40${\%} of $\log(2)$.\footnote{Bernhard
{\it et al.}\cite{bcl02} used  a mean-field decoupling  on correlation
functions to  compute the specific  heat of  this system. Their result
are  qualitatively similar  and  also shows  a  large  entropy deficit
($46${\%}  of  $\log(2)$).}  However this  low-temperature peak should
``contain'' almost no  energy (2\%), which  means that such peak would
have to  occur at very  low  temperatures.  In  order to estimate this
temperature, one  can  add a $\delta$-function peak  to  the curves of
Fig.~\ref{fig:PadeCV}  in  order  to recover  the  correct  energy and
entropy variations.  This   constrains both the location  (temperature
$T_1$) and the  weight of the  $\delta$-peak.   By averaging over  the
different curves of Fig.~\ref{fig:PadeCV}  one finds $T_1\simeq  0.05$
(resp.  $T_1\simeq 0.08$)   for  a ground-state   energy  $e_0=-0.865$
(resp.  $e_0=-0.875$).  These estimates  are   in agreement with   the
conclusions of the more elaborate treatment described below.

\section{Entropy method}

In this section we briefly summarize the method we  use to compute the
specific heat.  More details  can be found  in Ref.~\onlinecite{bm01}.
The  specific heat $c_v$ and the  temperature $T$ can be obtained from
the  entropy  $s$ as a  function of  the energy  $e$  using
basic thermodynamic relations:
\begin{eqnarray}
    T(e)&=&1/s'(e) \label{eq:Defbeta} \\
    c_v(e)&=&-\frac{s'(e)^2}{s''(e)} \label{eq:DefCv}
\end{eqnarray}

\begin{figure}
\begin{center}
\includegraphics[width=4.4cm]{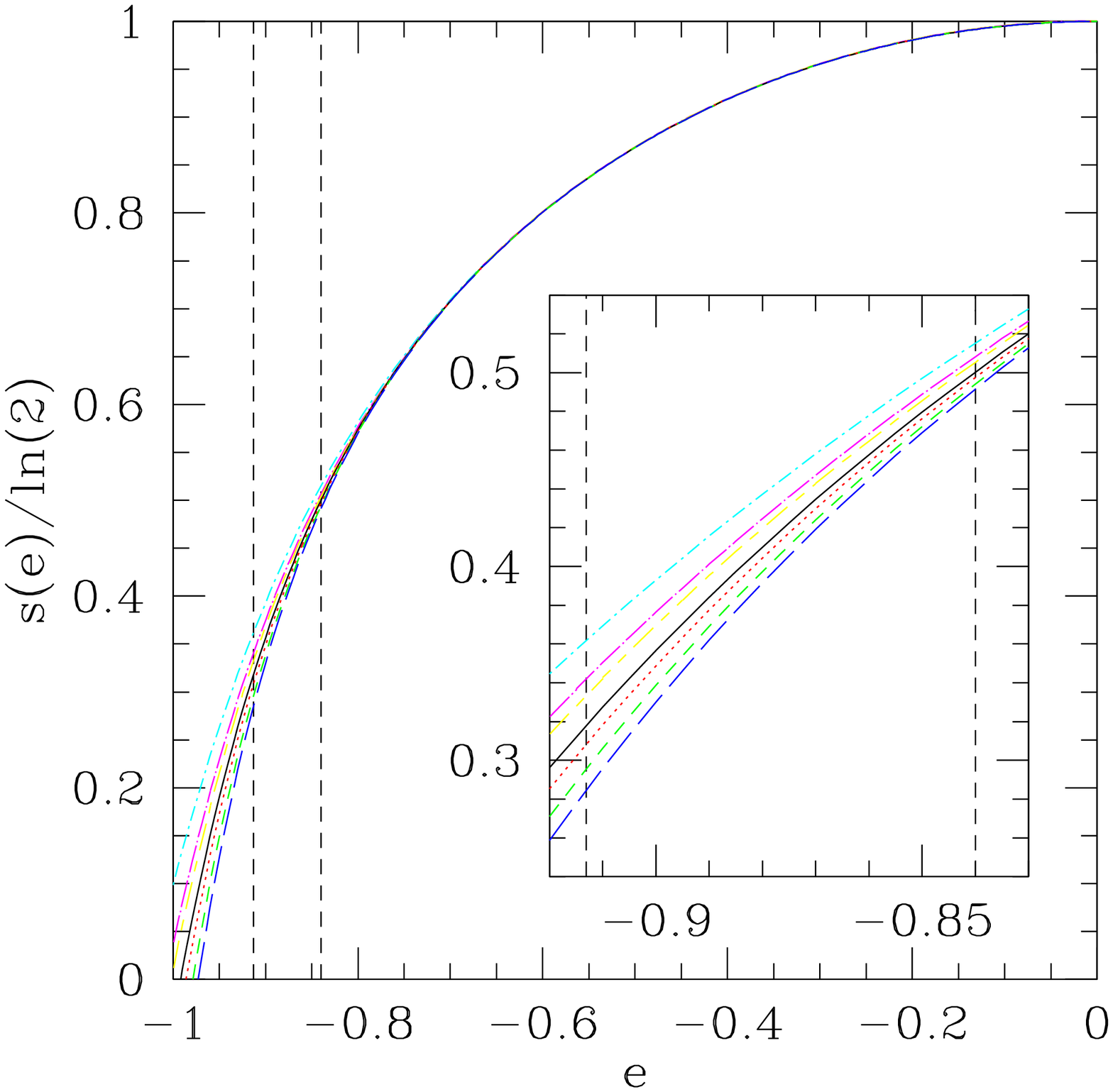}\hspace*{-0.89cm}
\includegraphics[width=4.4cm]{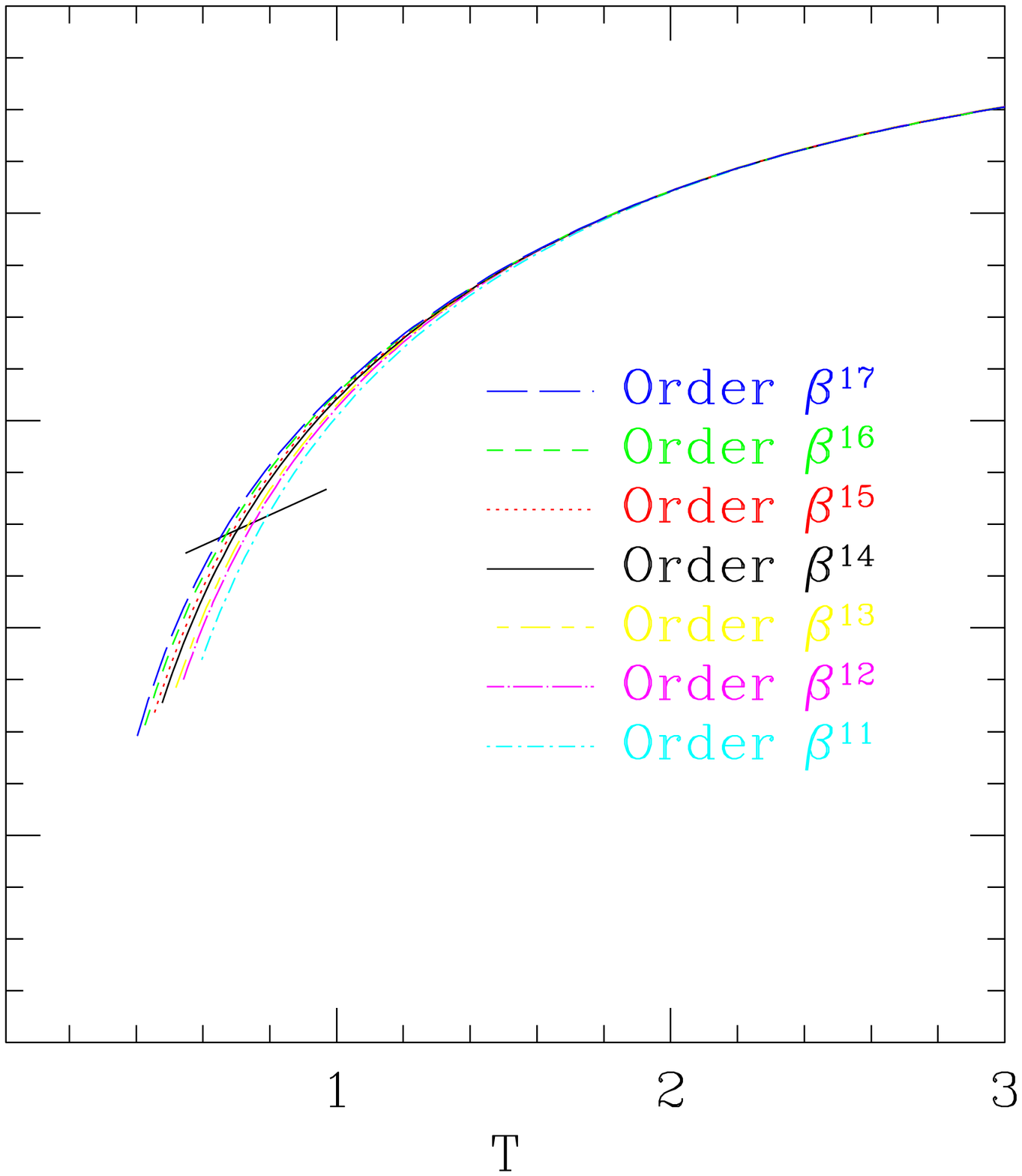}
\end{center}
\caption{(color online)
Left: Series expansion for $s(e)$ (entropy per  site) as a function of
the energy per site $e$.  The result of  the bare series are displayed
for   orders from $\beta^{11}$  to $\beta^{17}$.   The dashed vertical
lines indicate upper ($e=-0.84267$) and lower ($e=-0.909952$) rigorous
bounds   on     the   ground-state  energy      in  the  thermodynamic
limit.\cite{bpm04} Right: Same  data as on the  left panel but plotted
as  a function of temperature  $T=1/s'(e)$.  For each curve the lowest
temperature  corresponds to  $e=-0.909952$  (lower bound).  The  black
segment corresponds to $e=-0.84267$ (upper bound). }
\label{fig:se}
\end{figure}

From Eq.~\ref{eq:DefCv}  one can convert\cite{bm01} a high-temperature
series for  $c_v(T\to \infty)$ into a series  for $s(e\to0)$ ($e=0$ at
$T=\infty$  for  the Hamiltonian  of  Eq.~\ref{eq:H}).  The  truncated
series  are plotted  in Fig.~\ref{fig:se}. Using Eq.~\ref{eq:Defbeta},
the  entropy can  be plotted  as a function   of temperature (right of
Fig.~\ref{fig:se}). A good convergence is  observed down to relatively
low  energies ($e\sim  -0.75$)  but the corresponding  entropy remains
very  large (more than 60\%   of $\ln(2)$), although the  ground-state
energy  is  not much lower (the  ground-state  energy lies between the
dashed vertical   lines  in  Fig.~\ref{fig:se}).   These  result   are
consistent with  a   direct  analysis  of  the  series   for  $c_v(T)$
(Fig.~\ref{fig:PadeCV}).   In addition,  it  appears that the ``true''
$s(e)$ must be bent downward below the  curves of the truncated series
(shown in  Fig.~\ref{fig:se}) between $e_0$ and  $\sim -0.75$ in order
to  reach  $s=0$ at $e=e_0$.  Due  to  Eq.~\ref{eq:DefCv}, this almost
certainly implies  a low-energy  (and therefore  low-temperature) {\em
peak} in   $c_v(T)$.  This  paper  makes  this   idea more precise  by
computing the specific heat obtained  by forcing the entropy to vanish
at $e=e_0$.

The  advantage  of working on  $s(e)$ rather  than  $c_v(T)$ is that
a two-point  Pad\'e interpolation  can  be used to  set the
ground-state energy and  the   total entropy  of  the system.
However,  $s(e)$ is singular at $e=e_0$ (since  $T=1/s'(e)\to0$ when
$e\to e_0$). For this reason one  cannot directly approximate $s(e)$
by a rational fraction (Pad\'e  approximant). If we assume that the
specific-heat behaves as
\begin{equation}
c_v\simeq (T/c_0)^\alpha \label{eq:c0}
\end{equation}
 at  low temperature ($c_0$  has the dimension   of an energy)
and $s(e_0)=0$, $s(e)$
 behaves as:
\begin{equation}
    s(e\to e_0)\simeq \frac{(\alpha+1)^{\frac{\alpha}{\alpha+1}}}{\alpha}
        \left(\frac{e-e_0}{c_0}\right)^{\frac{\alpha}{\alpha+1}}
            \label{eq:se}
\end{equation}
The quantity
\begin{equation}
    G(e)=\frac{s(e)^{1+1/\alpha}}{e-e_0}
    \label{eq:G}
\end{equation}
is then  non-singular at $e=e_0$  and can be  approximated by a
Pad\'e form.\footnote{Eq.~\ref{eq:G} is slightly  different from the
original choice made  in Ref.~\onlinecite{bm01}.} The series for
$s(e\to0)$ must therefore be  transformed  into a  series for
$G(e\to0)$   before Pad\'e approximants can then  be  computed in
the   usual way. In what follows all the Pad\'e approximants will be
approximations to this function $G(e)$. If no finite-temperature
phase transition is expected, all approximants where  $G(e)$ has  a
pole or a zero, or where $s'(e)$ or $s''(e)$ vanishes  somewhere in
the interval $]e_0,0[$ must  be discarded. The remaining ones are
called ``physical'' for brevity.

\subsection{Low-temperature behavior of $c_v(T)$}

Unlike some simpler  magnets where the nature  of the ground-state and
elementary excitations  is known,\cite{bm01}  the qualitative behavior
of  the specific heat when $T\to0$  is unknown,  although a $\sim T^2$
scenario has been   proposed.\cite{ThermoKago00,bcl02} However, one  of  the
striking facts about the model is the unusually high density of states
{\em immediately above}  the ground-state.\cite{web98} From this it is
natural  to   expect gapless   elementary excitations.   If   we assume
quasi-particles with  a dispersion relation $\epsilon_k\sim k^{\gamma}$
we get a specific  heat $c_v\sim  T^\alpha$ with $\alpha=D/\gamma$  in
space  dimension    $D$.  The   (many   body) density  of    states is
$\rho(E_0+W)\sim
\exp{\left[N(W/N)^{\alpha/(\alpha+1)}\right]}$ where $N$ is the
system size (consequence of Eq.~\ref{eq:se}  with $e-e_0=W/N$). For an
energy $W$ of  order one above the ground-state,  a  density of states
$\rho\sim  1.15^{N}$   was  observed in  exact spectra    up to $N=36$
sites.\cite{web98} If this indeed holds up to the thermodynamic limit,
it would imply $\alpha=0$  ($\gamma=\infty$) and an  extensive entropy
at  zero   temperature.     This  is    unlikely     in  the   present
model\footnote{See Ref.~\onlinecite{msp03} for  a  related model where
this property is explicitly demonstrated. This, however, requires some
fine  tuning of the Hamiltonian  and there is  no reason to think that
the kagome Heisenberg model may have such  very special property. From
finite-size   spectra it  can however   be   difficult to resolve  the
difference between  a model with an  extensive entropy at $T=0$  and a
system with vanishing entropy   at $T=0$ but a  large  low-temperature
peak  in the  specific heat for  $T\ll J$.   We stress  that the $T=0$
entropy that follows from $\rho\sim 1.15^{N}$  is consistent with that
of  the low-temperature peak predicted  by  the present method:  about
20\%-25\%  of  $\log(2)$.}  but this  result points   to a rather flat
dispersion relation of the excitations,  probably with $\gamma>1$.  In
the following we will  consider the two cases  $\gamma=1$ ($\alpha=2$)
and  $\gamma=2$  ($\alpha=1$).\footnote{The  possibility  of  having a
gapped spectrum with  a  (very) small gap  cannot be  excluded.   This
would be the case if the system realizes a $\mathbb{Z}_2$ spin-liquid,
as  predicted in    large-$N$  approaches\cite{sachdev92}  or  if  the
ground-state        has    some         valence-bond       long-ranged
order.\cite{mz91,ns03,ba04}  A  thermally activated  behavior $c_v\sim
\exp{(-\Delta/T)}$ can be treated within the entropy method\cite{bm01}
and we applied it  to the  present  model.  The  gaps we obtained  are
rather large (of order 1) but the  specific-heat curves also exhibited
a low-temperature peak consistent with  the $c_v\sim T^\alpha$ results
discussed below.}

\subsection{Ground-state energy   and  convergence    of  the  different
Pad\'e approximants}

In   principle,  the  method  above  requires  the   knowledge of  the
ground-state energy $e_0$.  If the  value of $e_0$  is exact we expect
the procedure  to converge to  the exact $c_v$  if the number of known
terms in  the HT series increases  to infinity.  This is  in agreement
with our experience on solvable models (such as the spin-$\frac{1}{2}$
XY chain for  instance\cite{bm01}) where  the  full series  as well as
$e_0$ are known exactly.  Inversely, wrong values of $e_0$ cannot lead
to any convergence as the limiting $c_v$  would have to satisfy the HT
series at all orders but would have a different energy sum rule.  As a
consequence, when $e_0$ differs from the true ground-state energy, the
physical   approximants gets  fewer  (and/or more  scattered) when the
order of expansion  gets larger.  Of  course, the smaller the error on
$e_0$ the  longer  series  is needed  to observe   this departure from
convergence. From this we assume that the existence of a larger number
of  physical approximant is   an  {\em  indication}  that  $e_0$  (and
$\alpha$)  is closer to the  exact value.  However,  because a limited
number of  terms  of  the  series are  known,  this   only provides  a
qualitative  information   and does not   allow  determine  the energy
completely.

From exact diagonalizations on systems with up  to 36 sites, $e_0$
was evaluated  by  Waldtmann    {\it      et  al.}\cite{web98} to be
$e_0=-0.865\pm0.015$                     (see also
Refs.~\onlinecite{ze90,fbg01,fb03}). Variational calculations as
well as rigorous    bounds  on $e_0$  will be     discussed  in a
separate paper.\cite{bpm04}$^,$\footnote{    The  expectation value
of  the Hamiltonian in any first-neighbor  valence-bond covering is
$e=-0.75$. By optimizing the   wave-function in  the vicinity of
each triangle without any dimer, Elser    obtained a variational
state with   energy $e=-0.8333$.    This     rigorous upper    bound
was     refined by Pierre\cite{pierre99} who used an improved
variational state to prove that $e_0\le-0.84267$.  And exact  lower
bound  can  be obtained  by remarking that the ground-state  energy
per triangle  in the kagome AF is necessarily larger  than   the
ground-state energy of   an isolated triangle. This gives
$e_0\ge-1$.   By extending this reasoning   to larger clusters we
can improve this  lower  bound.  We can prove that $e_0\ge-0.909952.$
by exactly diagonalizing  a 24-site cluster with open boundary
conditions. Farnell, {\it  et  al.}  used a coupled cluster method
and  predicted   $e_0\simeq-0.8504$.\cite{fbg01} They recently
improved  their   calculation\cite{fb03}    and  obtained $e_0\simeq
-0.86208$   by fitting their   $n^{\rm  th}$-order results  by
$1/n^2$ corrections. We observed that their  bare data are better
described by $1/n$ corrections. In that case a fit gives $e_0\simeq
-0.875$.}

\begin{figure}
\begin{center}
\includegraphics[width=5.2cm]{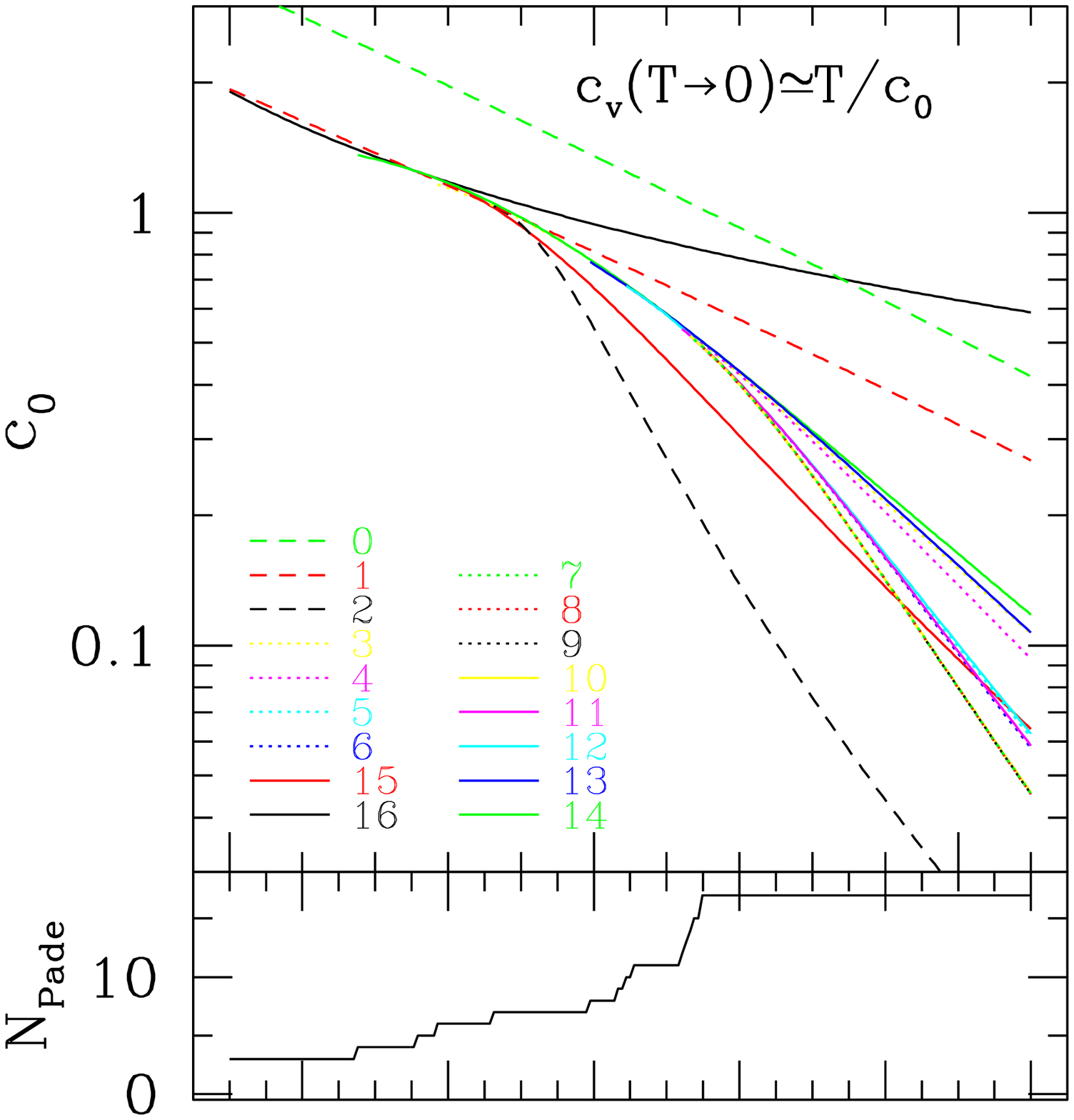}\hspace*{-1.63cm}
\includegraphics[width=5.2cm]{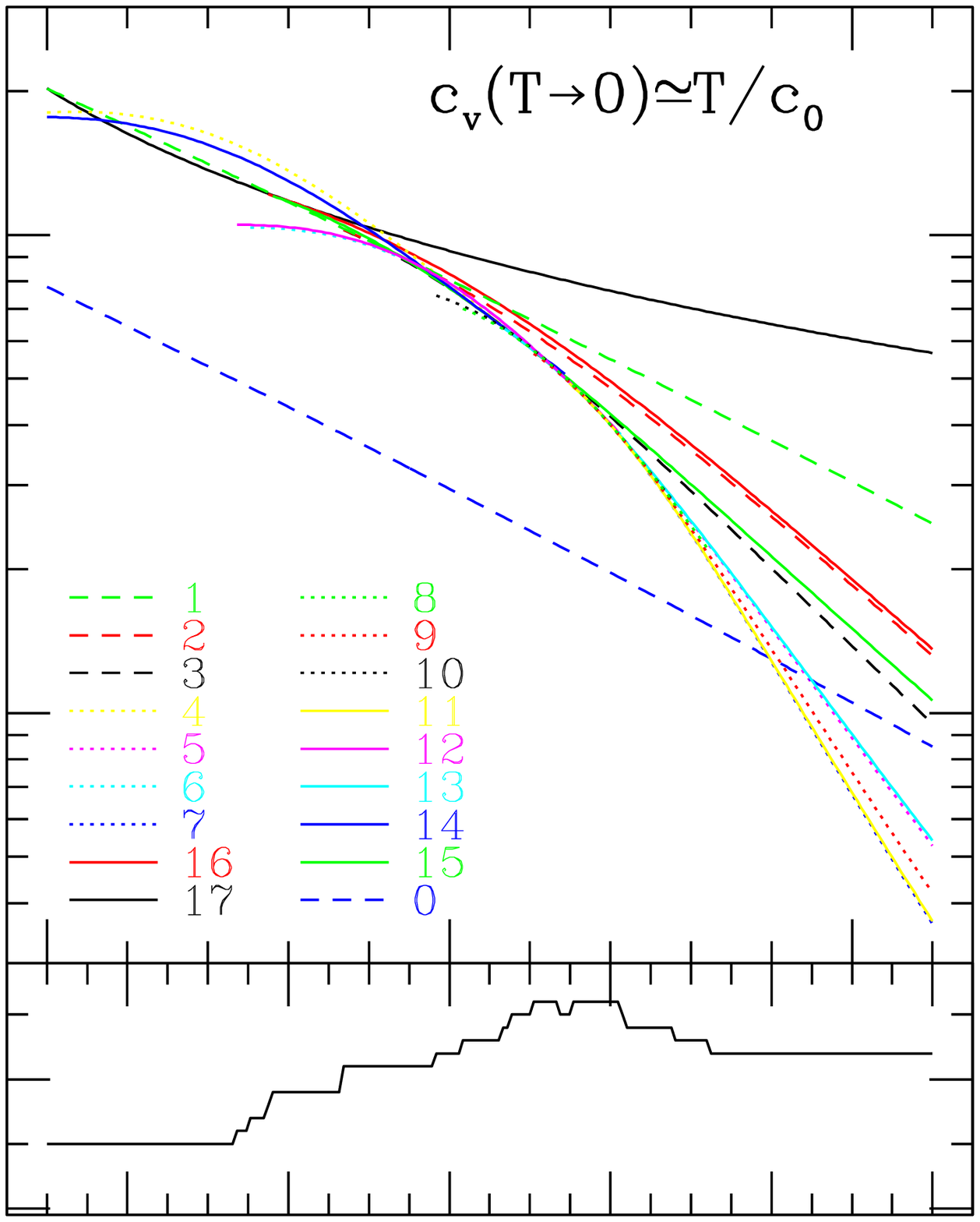}\\
\vspace*{-0.55cm}\includegraphics[width=5.2cm]{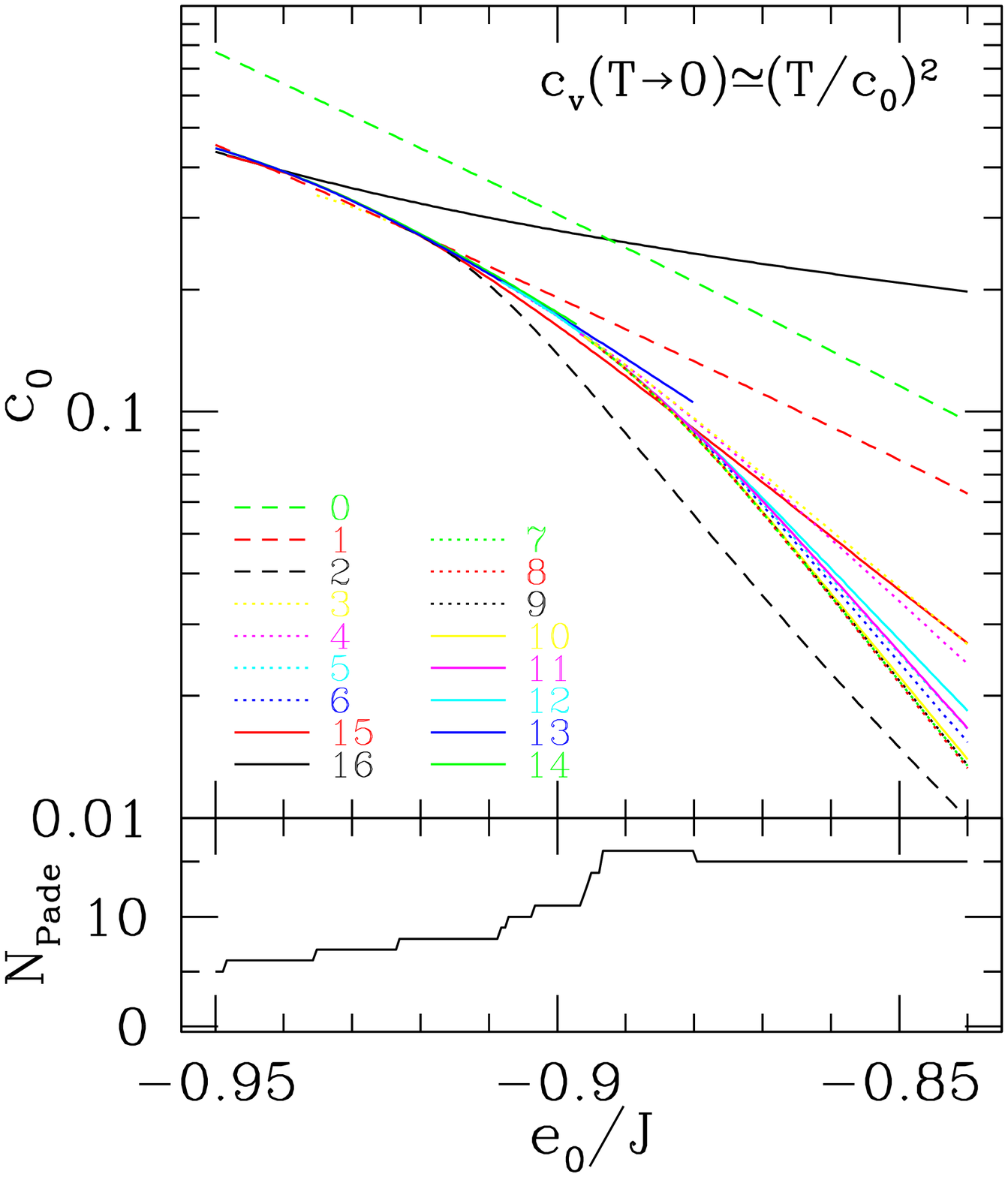}\hspace*{-1.63cm}
\includegraphics[width=5.2cm]{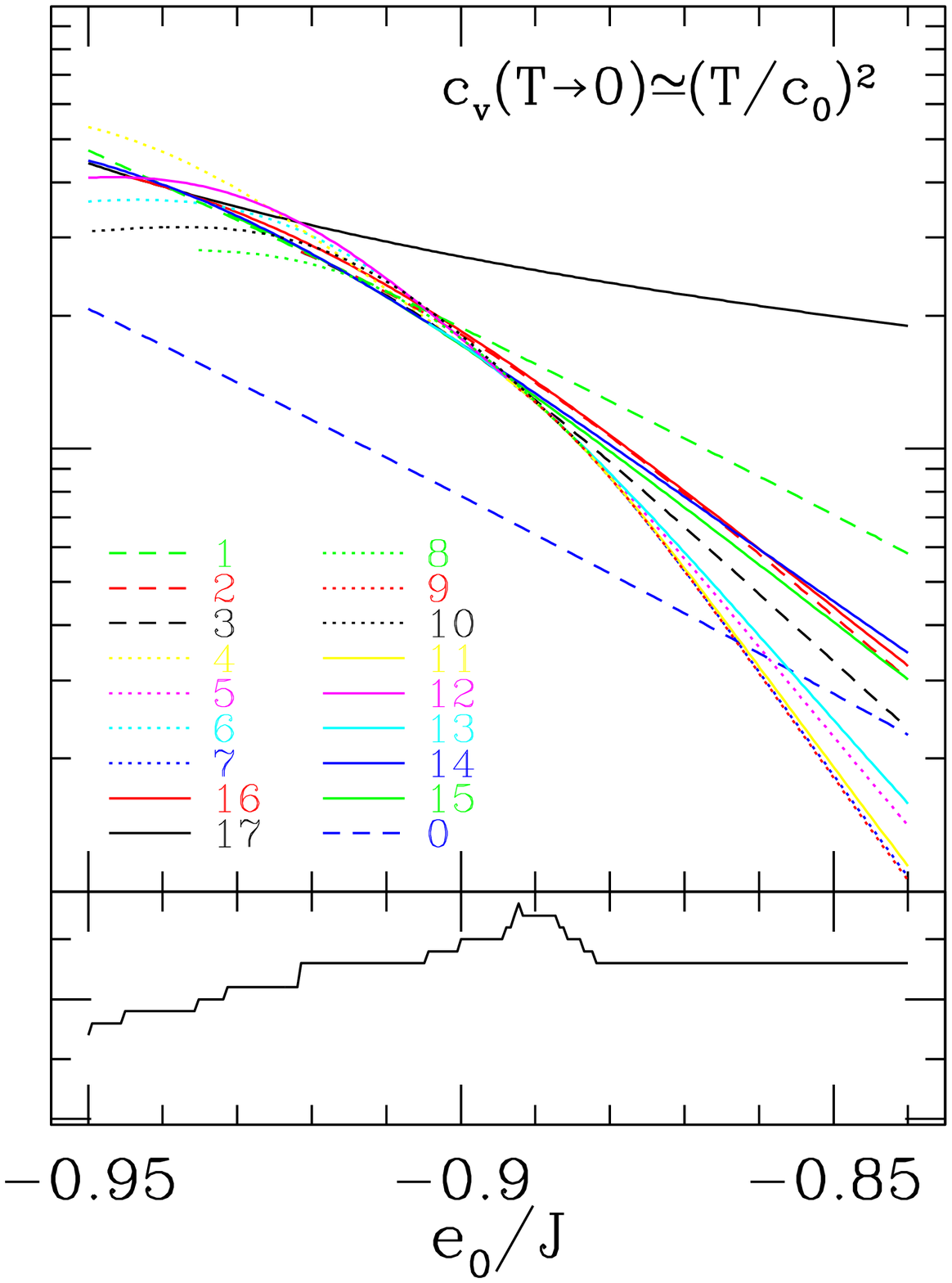}
\caption{
(color online)  Zero temperature limit  of $c_0$ (see Eq.~\ref{eq:c0})
for  the    different  Pad\'e approximants  as      a function of  the
ground-state energy.  Top: a $c_v\sim(T/c_0)$  behavior is assumed  at
low temperature.  Bottom:  $c_v\sim(T/c_0)^2$.  The degree $u$  of the
numerator of each approximant is given.  The degree of the denominator
is $d=n-u$  where   $n$ is the  order  of  the  series.   Left:  Order
$n=16$. Right:   Order $n=17$.   The  number of   physical approximant
($N_{\rm  Pade}$) is  plotted as  a function of   $e_0$  in each lower
panel.}
\label{fig:c0}
\end{center}
\end{figure}

The specific heat   curve can be  rather sensitive  to the choice of
$e_0$.  Since $e_0$ is not exactly known, it is important to perform
scans in order to see how the  specific-heat curve depends on $e_0$.
We observe that, for some choice of $e_0$ many Pad\'e approximants
at a given order give almost the same specific heat curve whereas
some other choice of $e_0$ leads to some significant scattering in
the specific heat curves. This can conveniently be seen, for
instance, by looking at the value of the different Pad\'e
approximants at $e=e_0$. Since $G(e_0)$ and $c_0$ (defined by
Eq.~\ref{eq:c0}) are simply related by
\begin{equation}
    G(e_0)=\frac{\alpha+1}{c_0\;\alpha^{1+1/\alpha}}
\end{equation}
we  plot  $c_0$ (which  has  a direct  physical   meaning in  term  of
$c_v(T\to 0)$)  in  Fig.~\ref{fig:c0} as a  function of  $e_0$ for all
physical Pad\'e  approximants at  order $\beta^{16}$ and  $\beta^{17}$
(both for $c_v(T)\sim T$ and $\sim T^2$  at low temperature). It turns
out that $c_0$  is representative of the  full  specific-heat curve in
the   sense that if two Pad\'e   approximants give ``close'' values of
$c_0$  (say   a relative   difference   less than   $10^{-3}$),  their
corresponding  specific-heat  curves  are  similar  (typical  relative
difference  of  $10^{-2}$)    {\em   for all   temperatures}.     This
low-temperature coefficient  $c_0$ is therefore   a useful quantity to
monitor how the $c_v(T)$ result depends on the choice of the degree of
the   Pad\'e   approximant.\footnote{The  physical   meaning  of  this
observation  is  that  if   two  specific heat   curves  have  i)  the
high-temperature expansion up to  some  relatively high order  ii) the
same  ground-state energy,  iii)  the same   entropy and iv)  the same
$c_v\sim(T/c_0)^\alpha$ limit  at  low temperature  they  must be very
``similar''.}  In all cases   the ``optimal'' energy region is  around
$e_0\simeq -0.88\pm0.02$.\footnote{For  lower energies the curves gets
closer  but the number of  physical Pad\'e  approximants actually gets
smaller  and smaller,   as  can  be   seen   in the  lower   panels of
Fig.~\ref{fig:c0}.}   We also observe a  gradual  shift of the optimal
region to higher energies as the order of the series is increased.  We
analyzed this effect   and  performed several  extrapolations to   the
infinite-order limit (data not shown).  It is not clear, however, that
this  indirect  method to  determine  the ground-state  energy is more
accurate than the other available estimates.\cite{bpm04}

\subsection{Low temperature peak in $c_v(T)$}
\begin{figure}
\begin{center}
\vspace*{0.5cm}
\includegraphics[width=4.4cm]{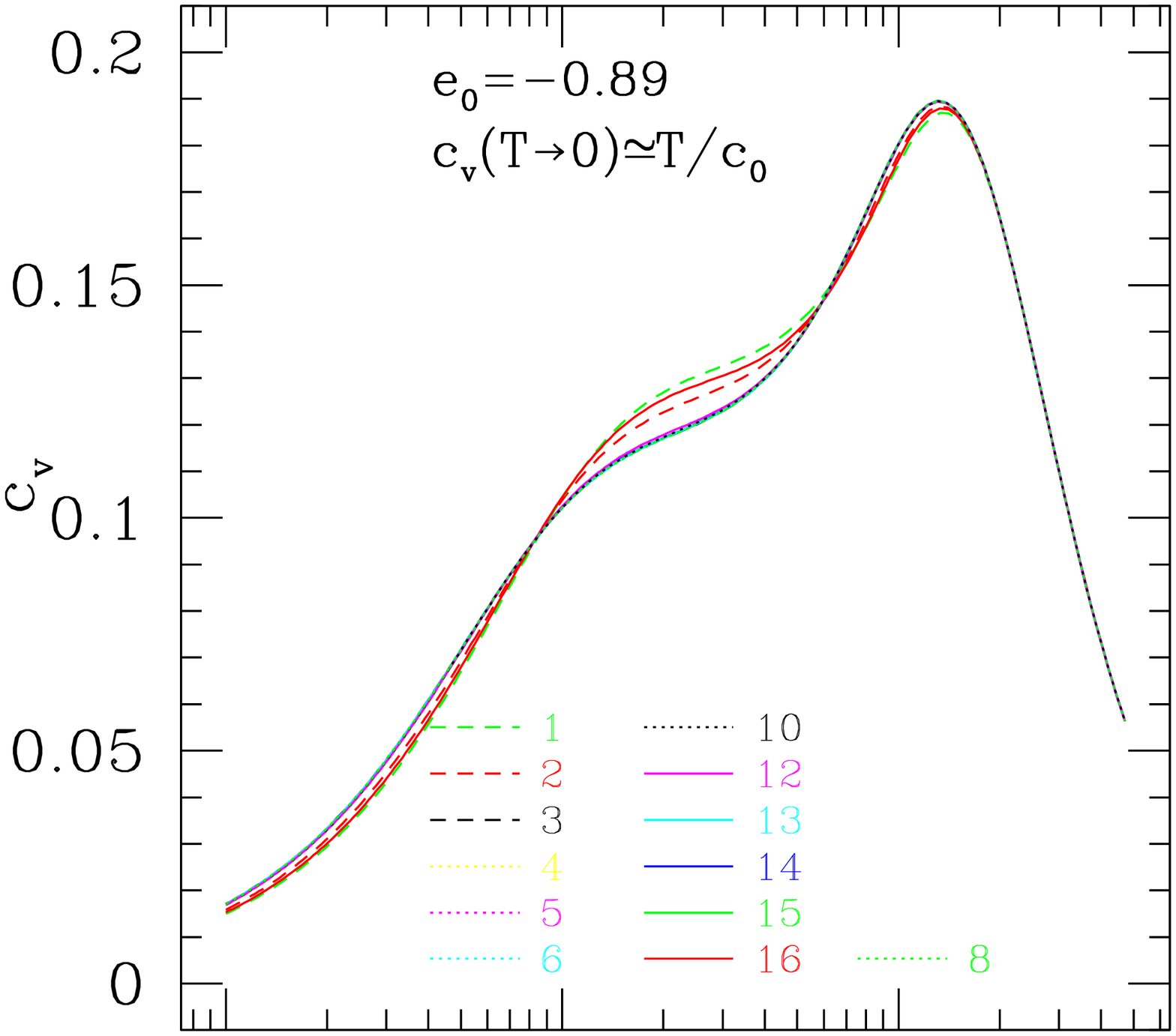}\hspace*{-0.79cm}
\includegraphics[width=4.4cm]{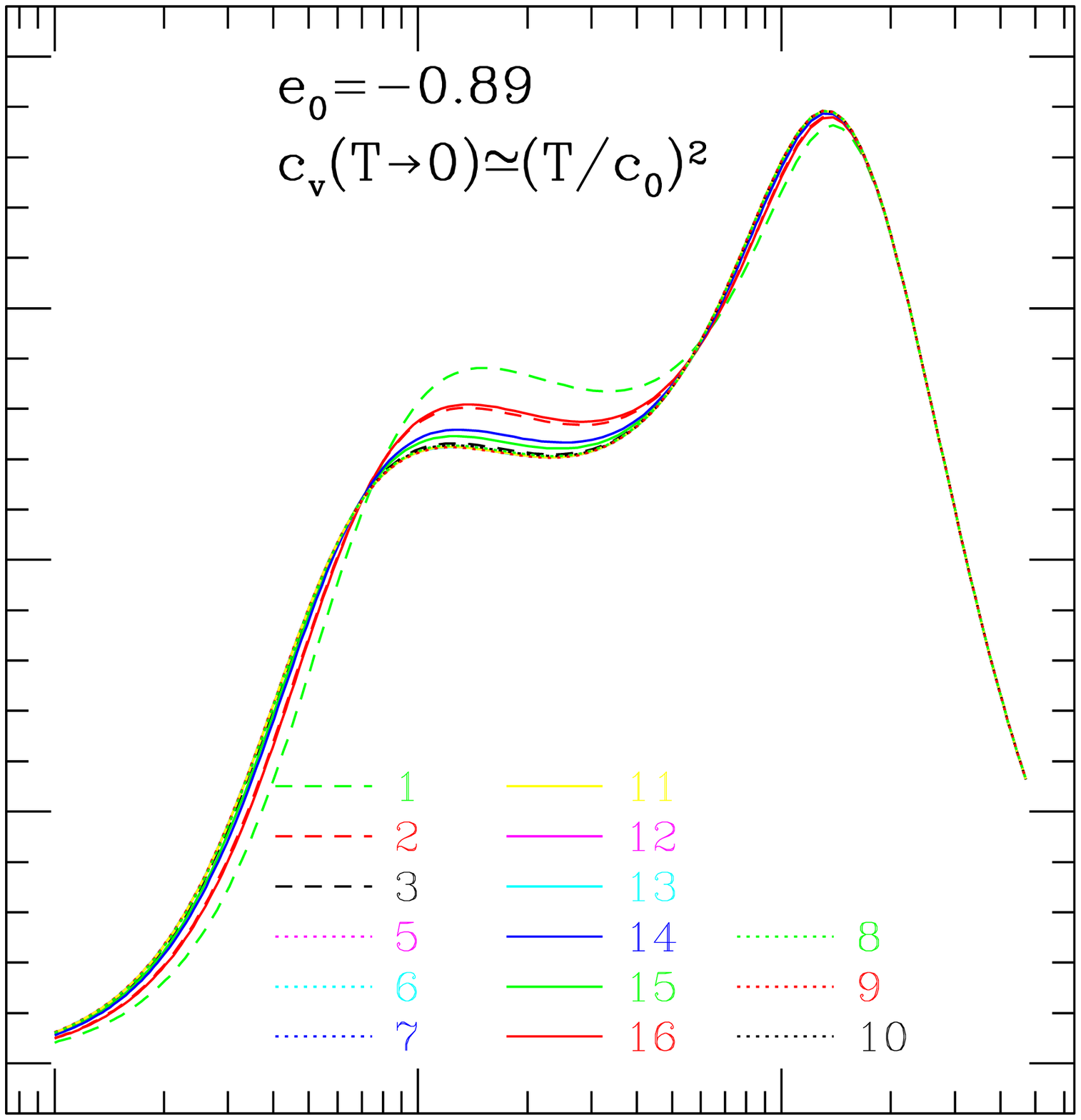}\\
\vspace*{-0.58cm}\includegraphics[width=4.4cm]{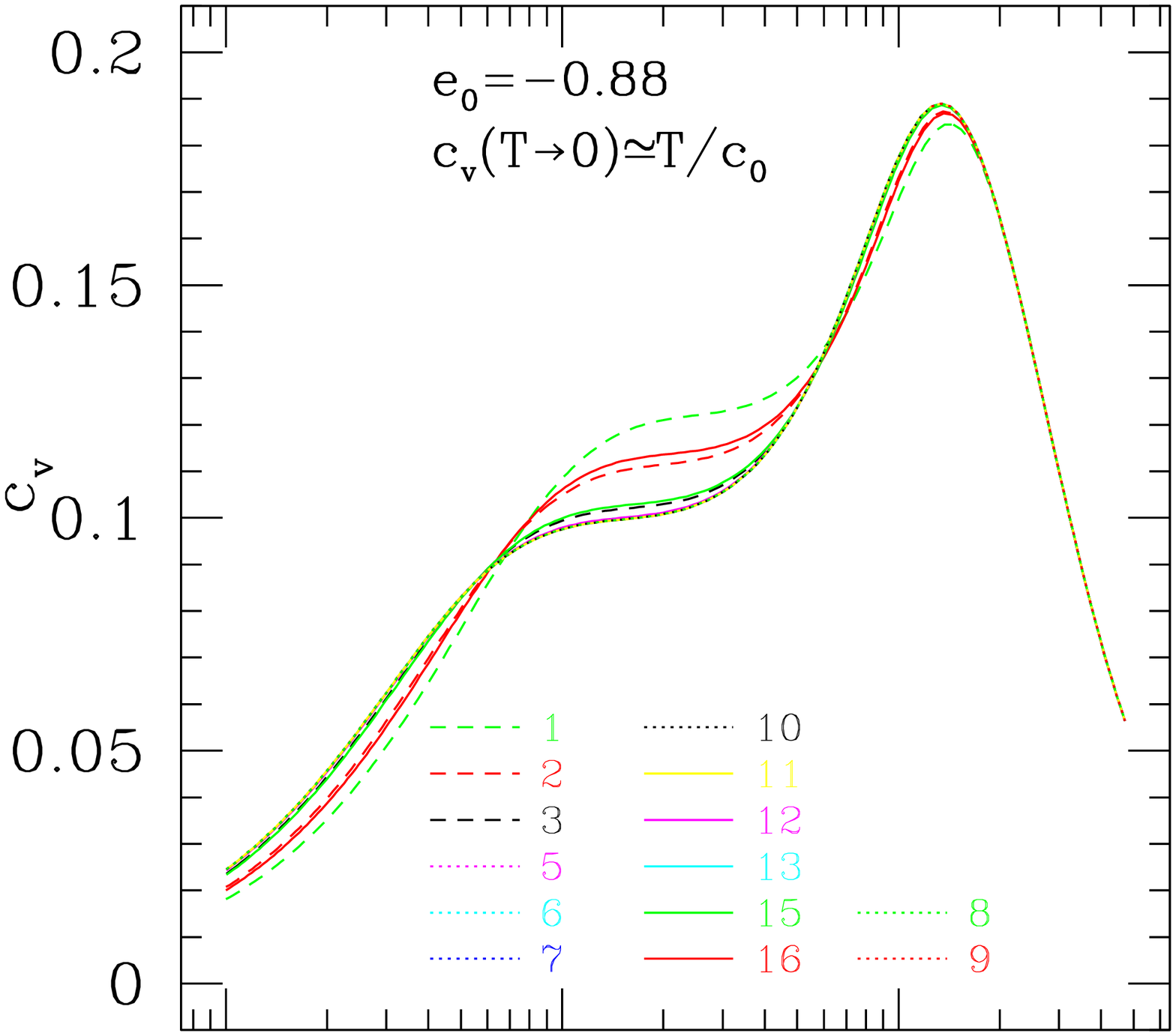}\hspace*{-0.79cm}
\includegraphics[width=4.4cm]{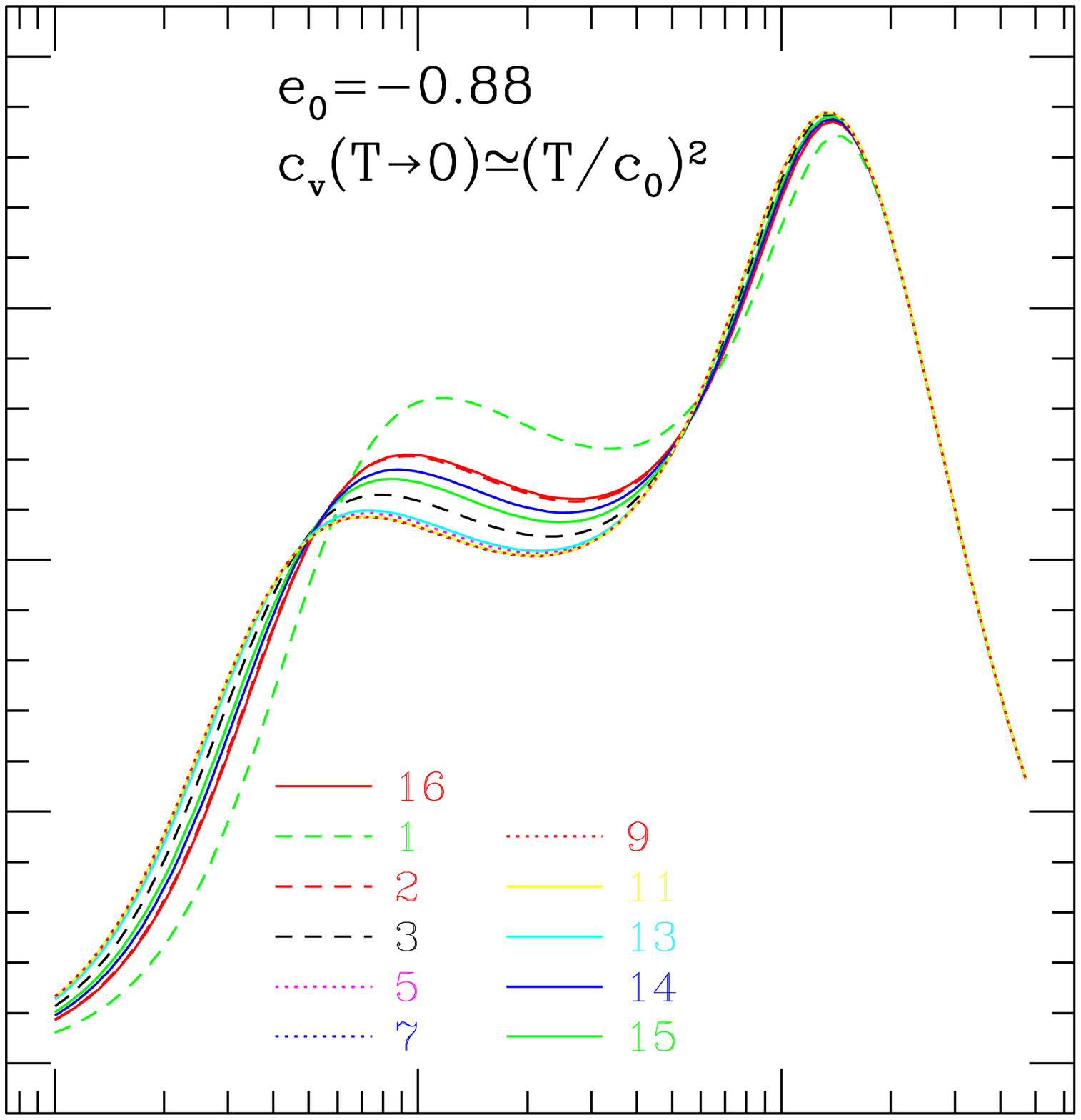}\\
\vspace*{-0.58cm}\includegraphics[width=4.4cm]{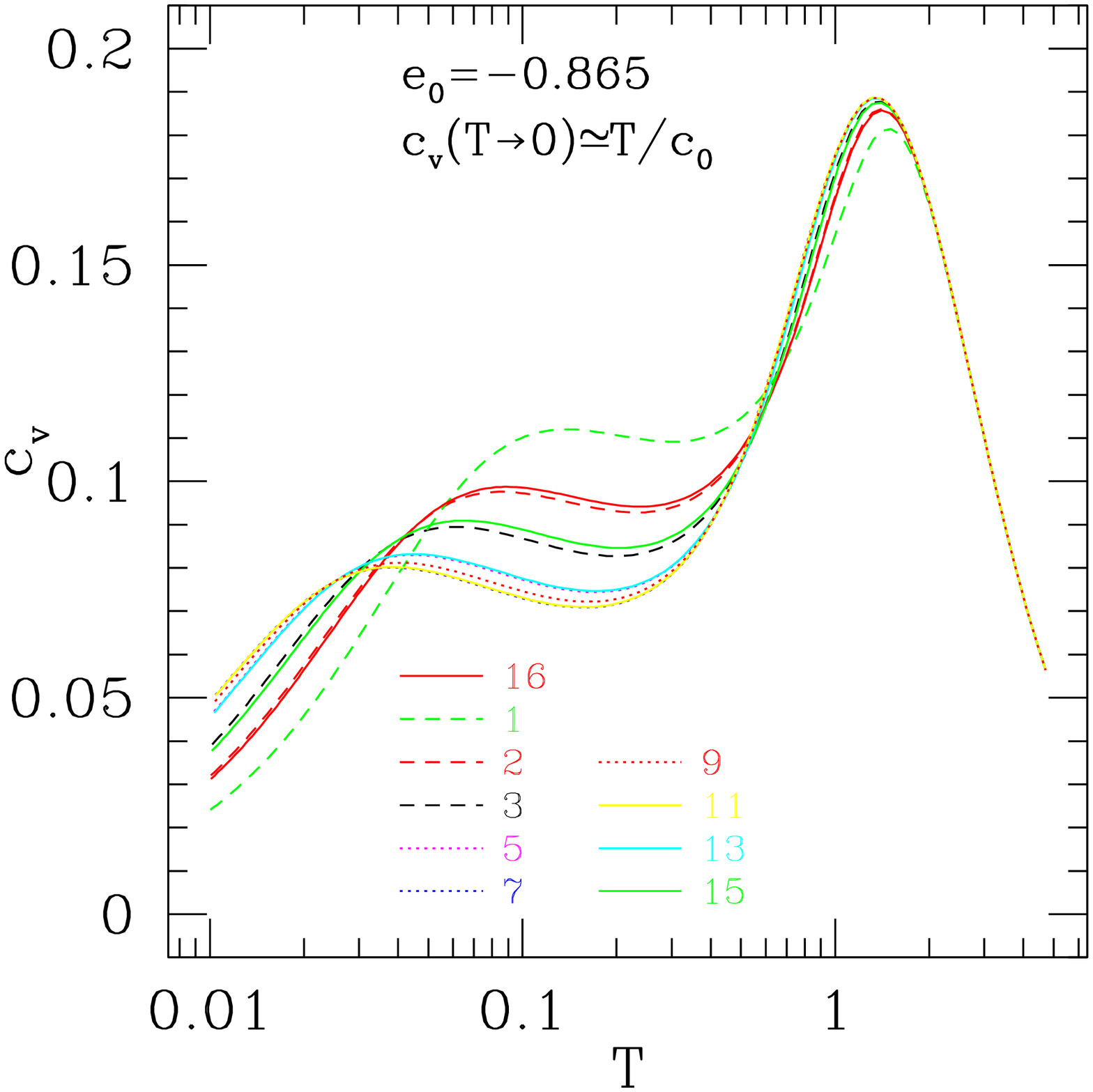}\hspace*{-0.79cm}
\includegraphics[width=4.4cm]{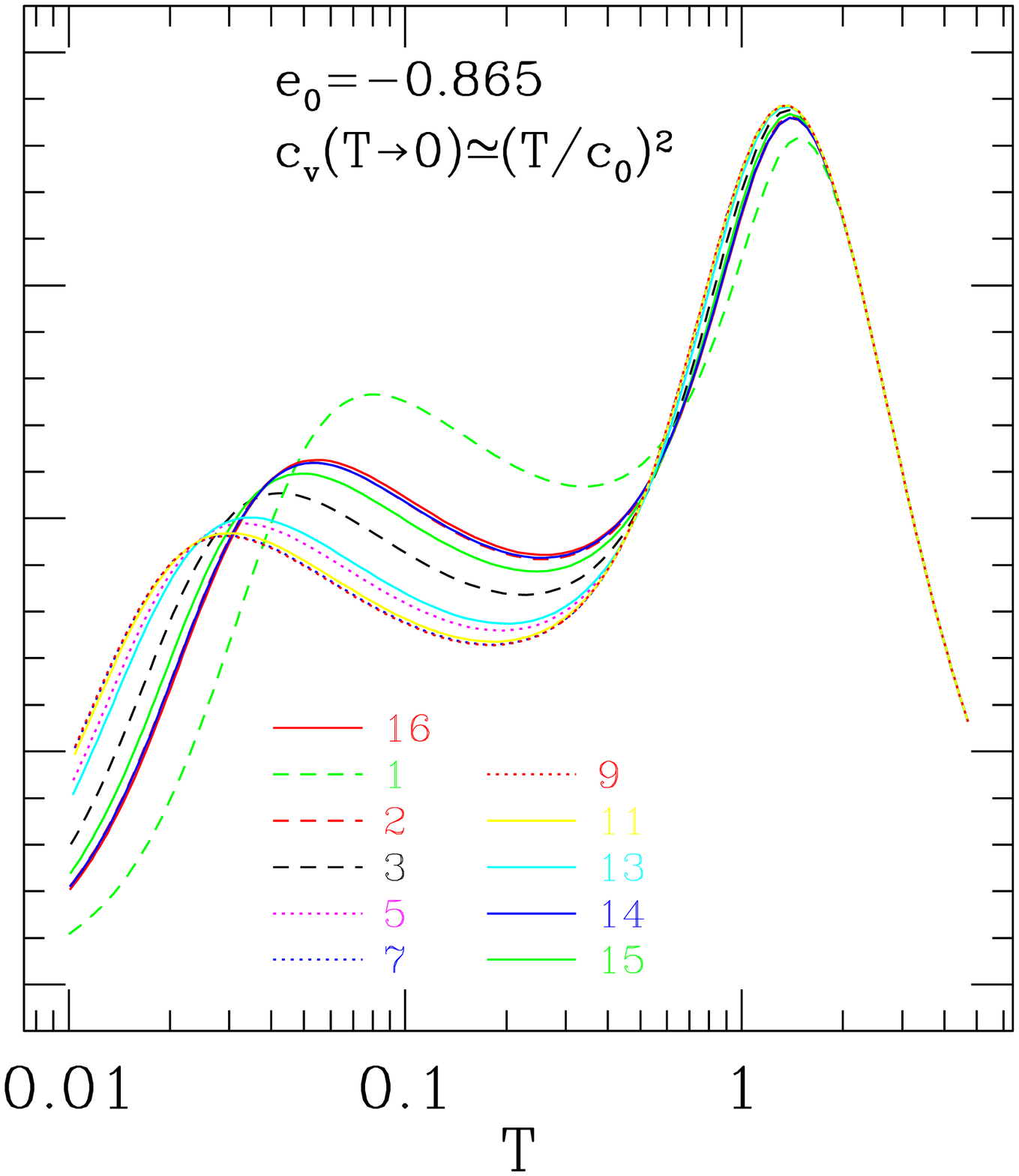}
\caption{
(color online) Specific  heat computed at  order $\beta^{17}$ with the
ground-state energies $e_0=-0.89$ (top),  $e_0=-0.88$ (center) and and
$e_0=-0.865$ (bottom).  Left: $c_v\sim T$.  Right: $c_v\sim T^2$.  The
different curves correspond to  all the physical Pad\'es approximants.
The degree $u$   of the numerator   is indicated (the  denominator has
degree $d=17-u$).}\label{fig:cv17}
\end{center}
\end{figure}

The curves    corresponding  to all  physical   approximants at  order
$\beta^{17}$ for  $e_0=-0.865$,  $e_0=-0.88$ and  $e_0=-0.89$, and for
$c_v\sim    T$     and     $c_v\sim      T^2$     are   shown       in
Fig.~\ref{fig:cv17}. Although some uncertainties remain concerning the
ground-state energy of  the   model  as well as   the  low-temperature
behavior of  the    specific heat, the  results   are  relatively well
converged down  to  $T\simeq  0.7$   and the   location  of  the  high
temperature peak is almost  independent  from the unknowns  ($e_0$ and
$\alpha$)   and         is     in  agreement        with      previous
studies.\cite{ey94,nm95,ThermoKago00,bcl02}  In  addition,  all    the
scenarios we  investigated give rise to a  low-temperature peak  (or a
shoulder) in the specific heat at $T\simeq 0.02 \sim 0.1$.

\begin{figure}
%\vspace{0.5cm}
\includegraphics[width=4.4cm]{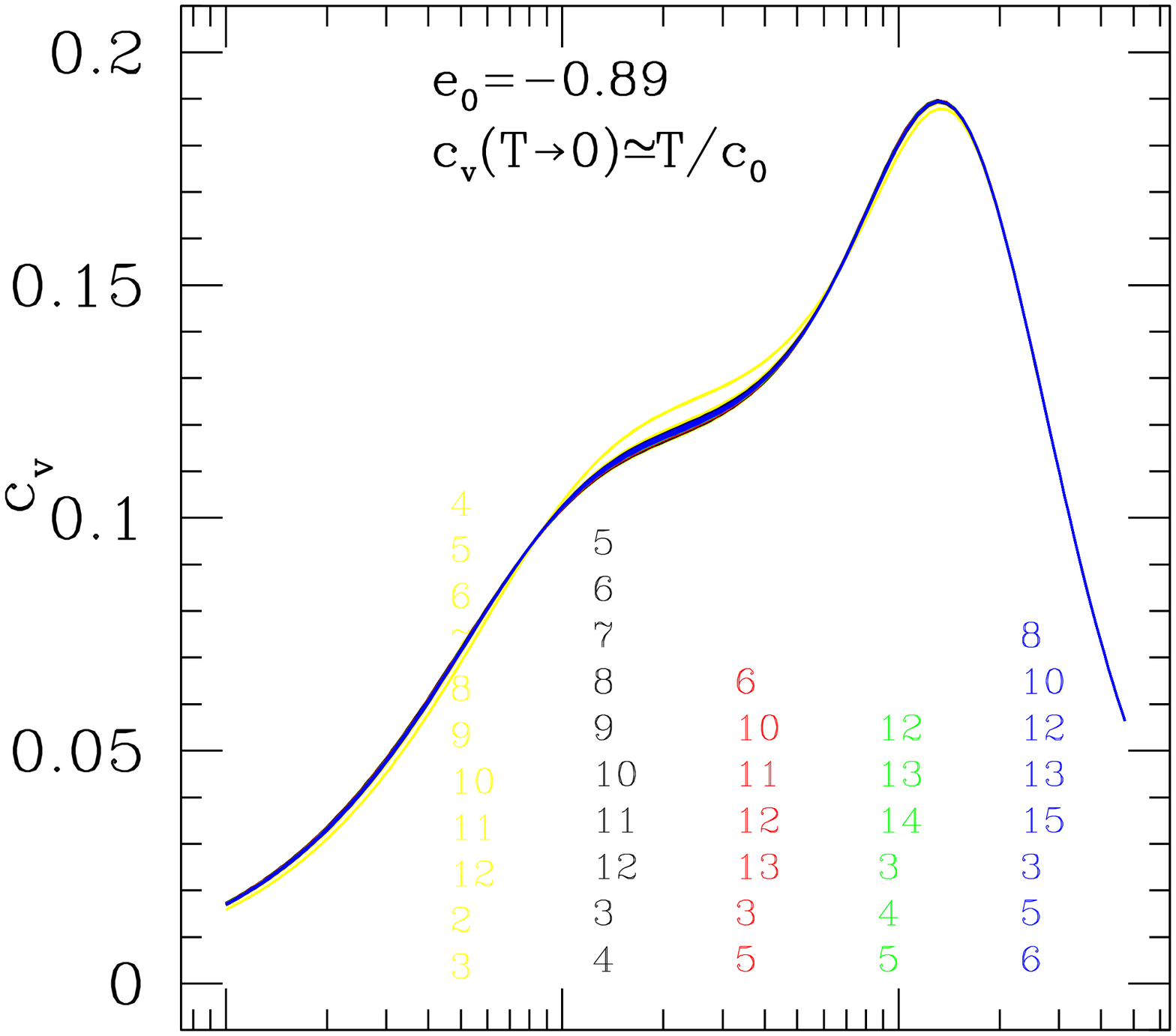}\hspace*{-0.79cm}
\includegraphics[width=4.4cm]{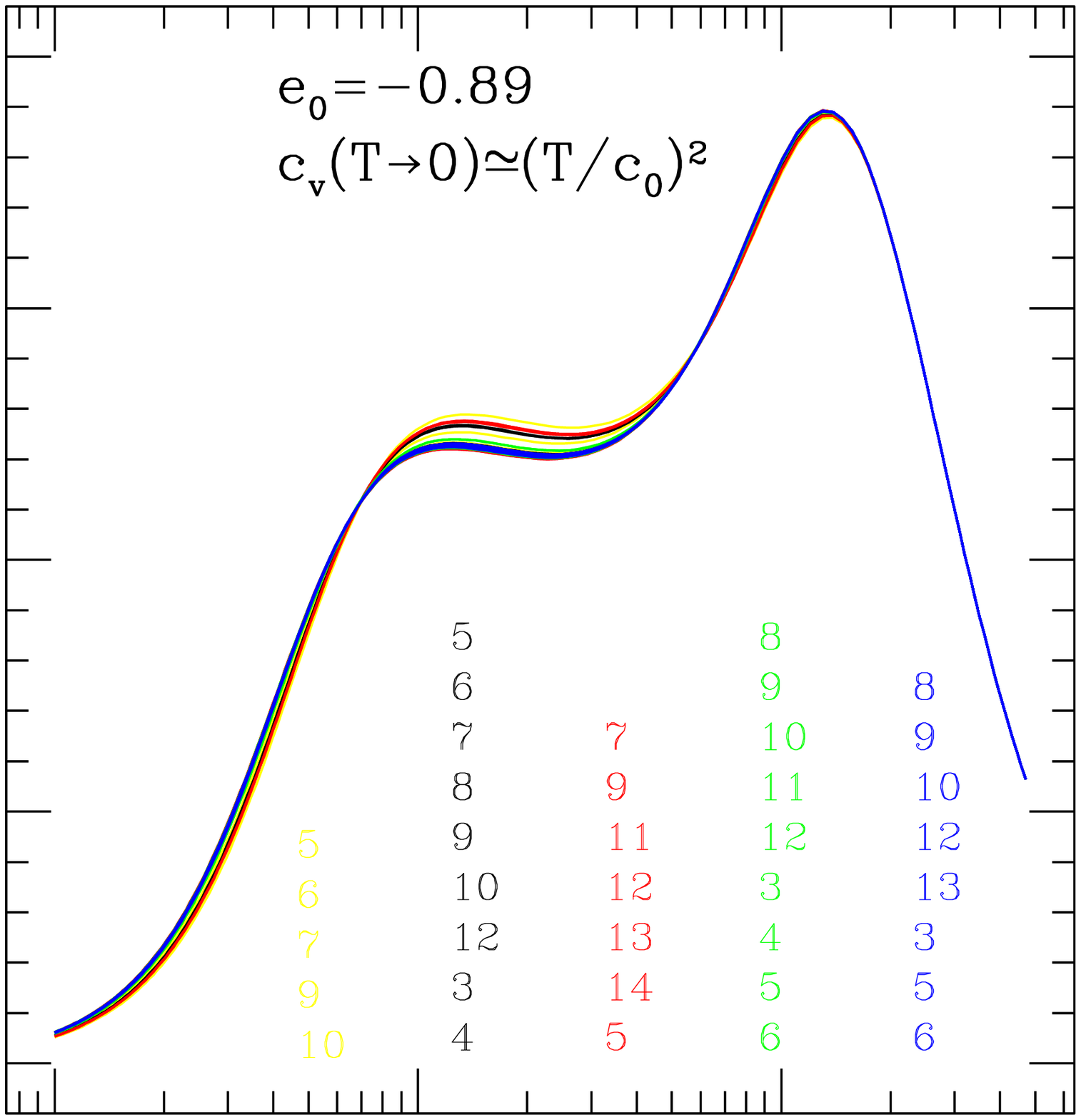}\\
\vspace*{-0.58cm}\includegraphics[width=4.4cm]{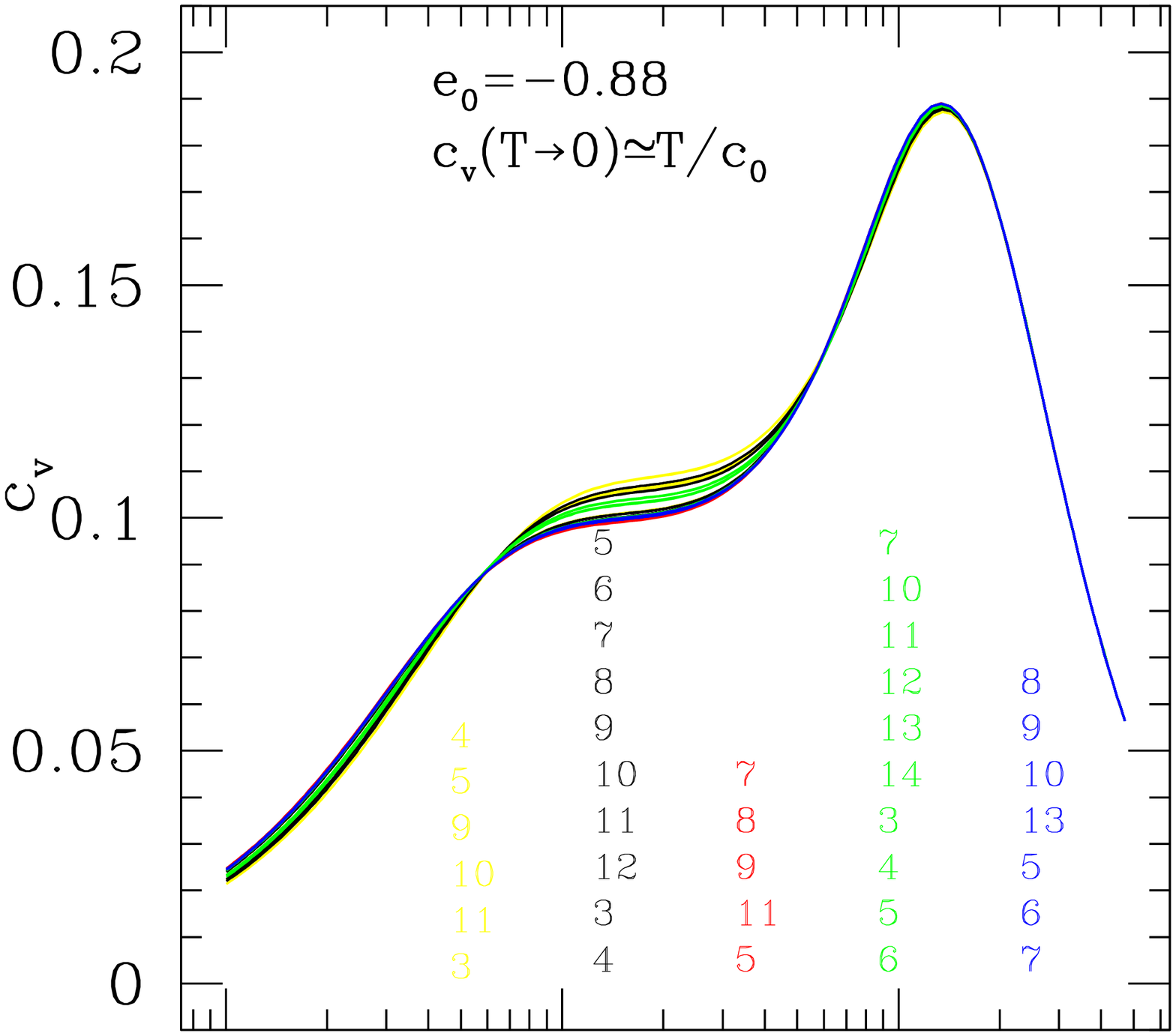}\hspace*{-0.79cm}
\includegraphics[width=4.4cm]{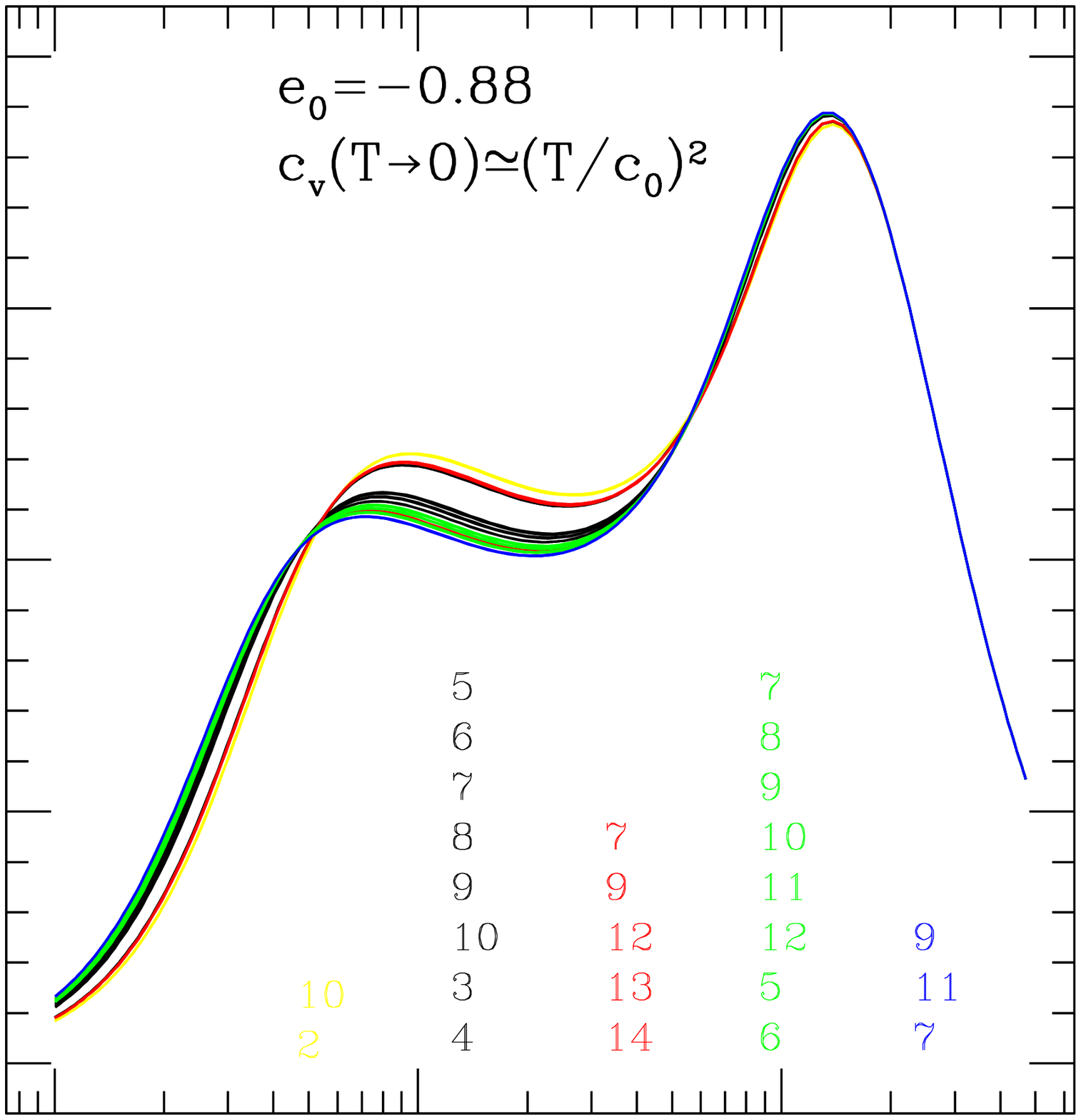}\\
\vspace*{-0.58cm}\includegraphics[width=4.4cm]{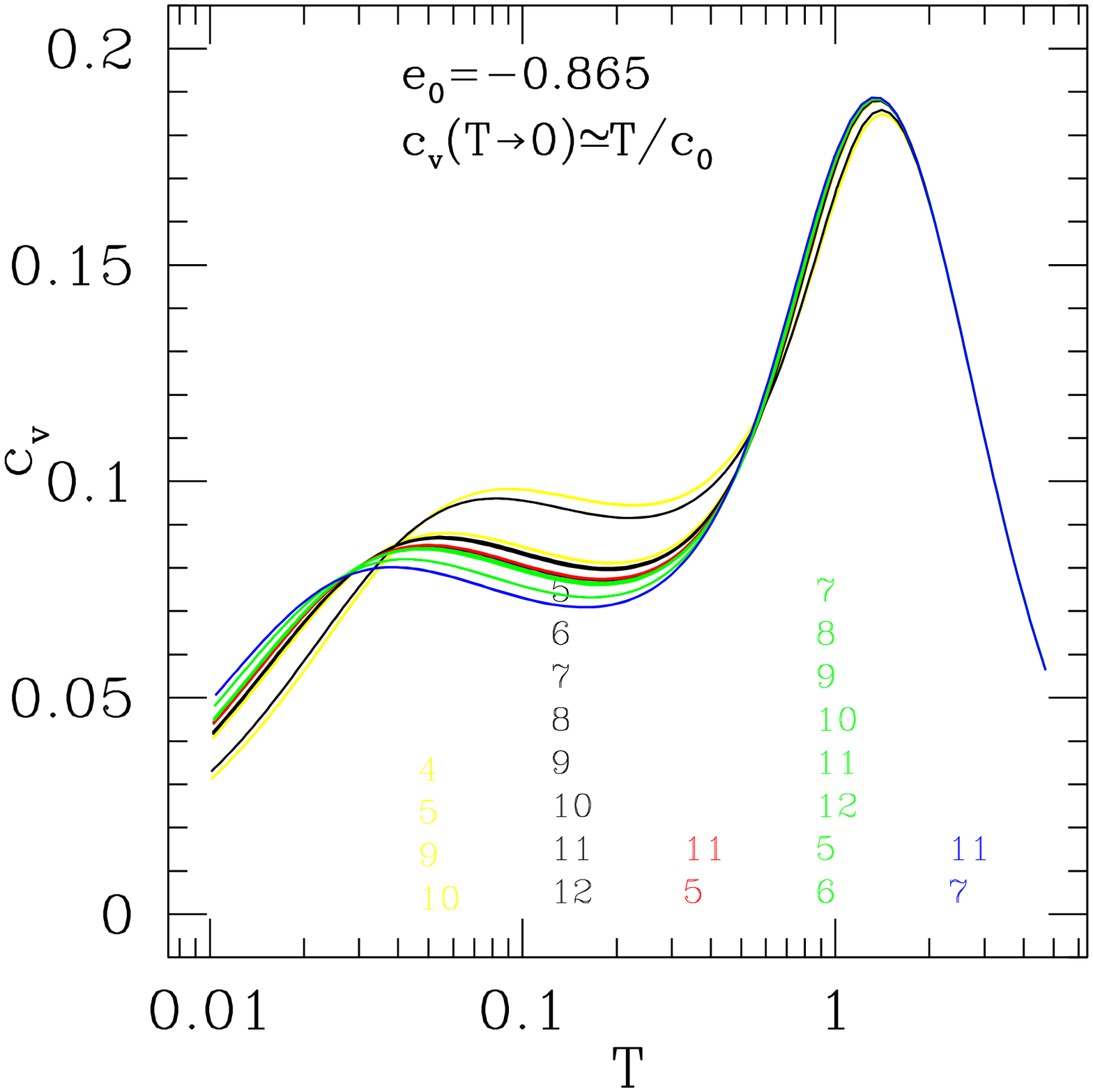}\hspace*{-0.79cm}
\includegraphics[width=4.4cm]{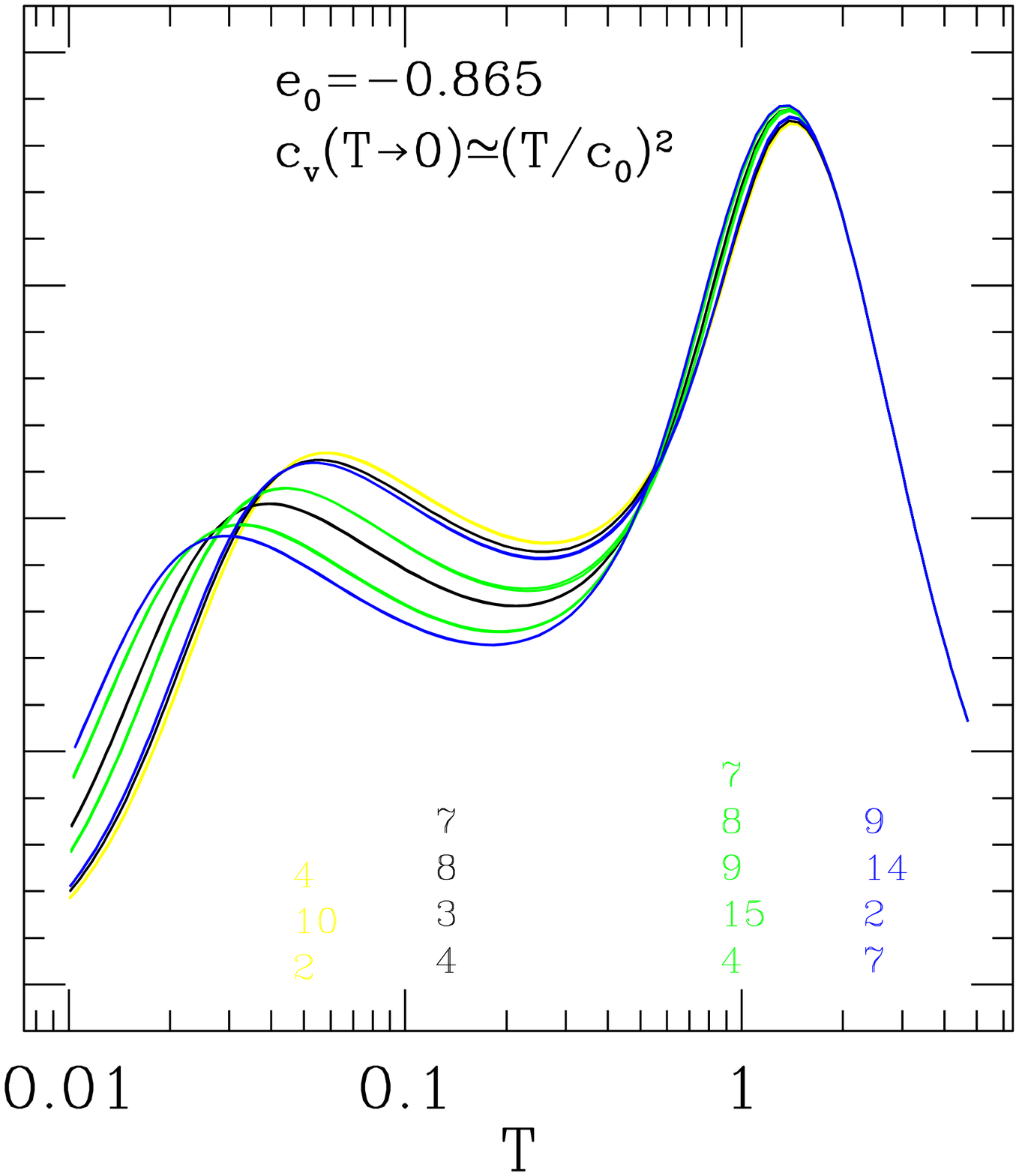}

\caption{(color online) Specific heat curves from order $\beta^{13}$
to   $\beta^{17}$ with  $e_0=-0.89$  (top),   $e_0=-0.88$ (center) and
$e_0=-0.865$   (bottom).   The degree    of   the  numerator of   each
approximant is indicated and each column corresponds  to a given order
($\beta^n$) of  the  series ($n=13,14,15,16$  and   $17$ from left  to
right).  Left  panels: $c_v\sim T$.  Right panels:  $c_v\sim T^2$.  In
all   cases  the specific heat  shows    a maximum around $T\simeq1.3$
(corresponding to  $e(T)\simeq-0.7$,  see Fig.~\ref{fig:se})   and   a
low-temperature peak (or shoulder). }\label{fig:cv}
\end{figure}

We  also  looked   at  the  order dependence  of   the   specific heat
curves. For a given value of the ground-state energy some approximants
give    similar curves     for   $c_v(T)$  while    some  other    are
``isolated''. The  later ones can be  recognized as isolated curves in
Fig.~\ref{fig:c0}.  According to  our experience\cite{bm01}  with this
method, those isolated approximants do  not reflect the convergence  to
the true function.  We  obtained  the results of Fig.~\ref{fig:cv}  by
keeping only the approximants  which value of  $c_0$  is at less  than
$3.10^{-3}$ from the $c_0'$ of another approximant. This selection was
repeated  from   orders $\beta^{13}$   to   $\beta^{17}$ for  the  six
combinations of ground-state  energies and  low-temperature  behaviors
used before.  As one can see, the low-temperature structure appears to
be a robust feature,  although a convergence of  the full curve is not
reached for $T\lesssim 0.6$. Still, a better convergence as a function
of  the order of  the series (and  a larger number  of physical Pad\'e
approximants) is  obtained  when   the  ground-state  energy  is   low
($e_0=-0.89$ or $e_0=-0.88$).  This suggests  that the actual value of
$e_0$ may be lower than $-0.865$, although $e_0=-0.89$ is probably too
low (compared to the available estimates\cite{web98,ze90,fbg01,fb03}).

For   $N=18$   spins,  exact    diagonalizations\cite{ey94}     gave a
low-temperature   peak  of the   specific  heat   at $T\simeq0.2$  and
$c_v\simeq0.17$.  An hybrid method\cite{ThermoKago00}  based on  exact
diagonalizations and high-temperature  series expansion gave a peak at
$T\simeq0.2$   and     $c_v\simeq0.17$    for   $N=36$   (see     also
Ref.~\onlinecite{nm95}).   Quantum  Monte-Carlo simulations for $N=72$
spins\cite{nm95} indicated that a  peak  may exist below  $T\simeq0.3$
for  this  system.  Those   results   obtained for small systems   are
qualitatively consistent with those of  Fig.~\ref{fig:cv} but our peak
is located  at a lower  temperature by at  least a factor  of two.  We
think that this discrepancy is likely to be due to finite-size effects
in previous studies.

\section{Conclusions}

By means of  a detailed high-temperature  series  analysis we provided
quantitative    estimates    for the   specific    heat  curve  of the
spin-$\frac{1}{2}$  Heisenberg  antiferromagnet on the kagome lattice.
Those results show a low-temperature peak in  the specific heat of the
model  for $T\lesssim 0.1$,   although its precise location cannot  be
determined   due to  uncertainties  on  the  ground-state energy.  The
corresponding degrees of  freedom are also  responsible  for the large
density  of singlet states  observed  in exact diagonalization studies
but their  nature, as well as the  nature of the  ground state itself,
remains to be explained.

\section{Acknowledgments}
It is a pleasure to  thank C.~Lhuillier, V.~Pasquier and P.~Sindzingre
for many  valuable discussions as well as  for  their collaboration on
related subjects.

\end{document}